
%
%
%
%
\input phyzzx
\tolerance=5000
\sequentialequations
\def\rl{\rightline}
\def\ll{\leftline}

\def\r#1{$\bf#1$}

\def\t1{{\tilde 1}}

\def\AEF{A.E.Faraggi}
\def\DVN{D.V.Nanopoulos}

\def\SSM{supersymmetric standard model}
\def\NPB#1#2#3{Nucl.Phys.B {\bf#1} (19#2) #3}
\def\PLB#1#2#3{Phys.Lett.B {\bf#1} (19#2) #3}
\def\PRD#1#2#3{Phys.Rev.D {\bf#1} (19#2) #3}
\def\PRL#1#2#3{Phys.Rev.Lett. {\bf#1} (19#2) #3}
\def\PRT#1#2#3{Phys.Rep. {\bf#1} (19#2) #3}

\def\vev#1{\left\langle #1\right\rangle}
\def\l{\langle}
\def\r{\rangle}

\REF\ALTARELLI{G. Altarelli, unpublished talk given at the conference
{\it Around the Dyson Sphere}, Princeton, NJ, April 8-9, 1994.}
\REF\HETE{D.J.Gross, J.A.Harvey, J.A.Martinec and R.Rohm,
                        \PRL{54}{85}{502}; \NPB{256}{86}{253}.}
\REF\CY{P. Candelas, G.T. Horowitz, A. Strominger and E. Witten,
\NPB{258}{85}{46}.}
\REF\DHVW{L.Dixon, J.A.Harvey, C.Vafa and E.Witten,
\NPB{274}{86}{285}.}
\REF\NARAIN{K.S. Narain, \PLB{169}{86}{41};
W. Lerche, D. L{\H u}st and A.N. Schellekens,
\NPB{287}{87}{477}.}
\REF\FFF{I.Antoniadis, C.Bachas, and C.Kounnas, \NPB{289}{87}{87};
H.Kawai, D.C.Lewellen, and S.H.-H.Tye, \NPB{288}{87}{1}.}
\REF\GEPNER{D. Gepner, \PLB{199}{87}{380};\NPB{296}{88}{57}.}
\REF\REVAMP{I. Antoniadis, J. Ellis, J. Hagelin, and \DVN, \PLB{231}{89}{65};
I. Antoniadis, G. K. Leontaris and J. Rizos, \PLB{245}{90}{161};
J. Lopez, D.V. Nanopoulos and K. Yuan, \NPB{399}{93}{654}, hep-th/9203025.}
\REF\SUTHREE{B. Greene {\it{el al.}},
Phys.Lett.{\bf B180} (1986) 69;
Nucl.Phys.{\bf B278} (1986) 667;  {\bf B292} (1987) 606;
R. Arnowitt and  P. Nath, Phys.Rev.{\bf D39} (1989) 2006; {\bf D42}
(1990) 2498; Phys.Rev.Lett. {\bf 62} (1989) 222.}
\REF\SSM{L.E. Iba{\~n}ez {\it{et al.}}, Phys.Lett.
{\bf B191}(1987) 282; A. Font {\it{et al.}},
Phys.Lett. {\bf B210} (1988)
101; A. Font {\it{et al.}},
Nucl.Phys. {\bf B331} (1990)
421; D. Bailin, A. Love and S. Thomas,
Phys.Lett.{\bf B194} (1987) 385;
Nucl.Phys.{\bf B298} (1988) 75; J.A. Casas, E.K. Katehou and C. Mu{\~n}oz,
Nucl.Phys.{\bf B317} (1989) 171.}
\REF\FNY{\AEF, D.V.Nanopoulos and K.Yuan, \NPB{335}{90}{347}.}
\REF\EU{\AEF,  \PLB{278}{92}{131}.}
\REF\TOP{\AEF, \PLB{274}{92}{47}.}
\REF\SLM{\AEF, \NPB{387}{92}{239}, hep-th/9208024.}
\REF\W{S. Weinberg, \PRD{26}{82}{475}; S. Sakai and T. Yanagida,
\NPB{197}{82}{533}; R. Arnowitt, A.H. Chamsdine and P. Nath,
\PRD{32}{85}{2348}.}
\REF\KLN{S.Kalara, J.Lopez and D.V.Nanopoulos, \NPB{353}{91}{650}.}
\REF\YUKAWA{\AEF, \PRD{47}{93}{5021}.}
\REF\FM{\AEF, \NPB{407}{93}{57}, hep-ph/9210256;
              \PLB{326}{94}{62}, hep-ph/9311312.}
\REF\DSW{M.Dine, N.Seiberg and E.Witten, Nucl.Phys.{\bf B289} (1987) 589.}
\REF\NRT{\AEF, \NPB{403}{93}{101}, hep-th/9208023.}
\REF\CKM{\AEF~ and E.Halyo, \PLB{307}{93}{305}, hep-ph/9301261;
          \NPB{416}{94}{63}, hep-ph/9306235.}
\REF\GLT{B. Grzadkowski, M. Lindner and S. Theisen,
\PLB{198}{87}{64}.}
\REF\GCU{\AEF, \PLB{302}{93}{202}, hep-ph/9301268.}
\REF\SUSYX{\AEF~ and E.Halyo, IASSNS--HEP--94/17, hep-ph/9405223.}
\REF\GPR{For a recent review see,
A. Giveon, M. Porrati and E. Rabinovici, RI--1--94, NYU--TH.94/01/01,
hep-th/9401139.}
\REF\FHn{\AEF~ and E. Halyo, \PLB{307}{93}{311}.}
\REF\SQCD{D. Amati {\it et. al.}, \PRT{162}{88}{169}.}
\REF\LT{D. L\"ust and T. Taylor, \PLB{253}{91}{335}.}
\REF\KLNI{S. Kalara, J.L. Lopez, \DVN, \PLB{275}{91}{304}, hep-ph/9110023.}

\singlespace
\rl{IASSNS--HEP--94/31}
\rl{May 1994}
\baselineskip=18pt
\bigskip
\centerline{\bf{Realistic Superstring Models
{\footnote*{\it Talk presented at the International Conference on Unified
Symmetry - in the Small and in the Large, Coral Gables, FL, 27-30 Jan 1994.
}}
}}

\bigskip
\centerline{Alon E. Faraggi
{\footnote\dagger{SSC fellow.}}{\footnote\ddagger{
                 e--mail address: faraggi@sns.ias.edu}}
}
\smallskip
\centerline {School of Natural Science, Institute for Advanced Study}
\centerline {Olden Lane, Princeton, NJ 08540}

\bigskip
\centerline{ABSTRACT}
\baselineskip=14pt
I discuss the construction of realistic superstring standard--like models
in the four dimensional free fermionic formulation.
I discuss the massless spectrum of the
superstring standard--like models and the
texture of fermion mass matrices.
These models suggest an explanation for the top quark mass hierarchy.
At the cubic level of the superpotential only the top quark get a mass term.
The lighter quarks and leptons obtain their mass terms from
nonrenormalizable terms that are suppressed relative to the cubic order term.
A numerical estimate yielded $m_t\sim175-180~GeV$.
The suppression of the lightest generation masses results from
the horizontal symmetries in the superstring models. The problems
of neutrino masses, gauge coupling unification and hierarchical SUSY
breaking are discussed. I argue that the realistic
features of these models are due to the underlying $Z_2\times Z_2$ orbifold,
with standard embedding,
at the free fermionic point in toroidal compactification space.

\baselineskip=18pt
\pagenumbers

\bigskip
\ll{\bf 1. Introduction}

The most fundamental problem in high energy physics is the nature of the
mechanism responsible for breaking the electroweak symmetry and for
generating the fermion masses. Two main school of thoughts were developed to
address this problem. The first assumes
that the origin of symmetry breaking is
dynamic and that the scalars doublets are composite. The second, assumes
the existence of fundamental scalar representations and tries to
incorporate the Standard Model into a fundamental theory, which unifies
the known interactions at a much higher scale.
In view of LEP precision data, the second approach is more
successful [\ALTARELLI].
Ultimately, future hadron colliders will determine the nature
of the electroweak symmetry breaking mechanism.

If we accept the notion of fundamental scalar representations and
unification, we must question what is the fundamental scale of unification.
Slow evolution of the Standard Model gauge couplings and
proton lifetime support a big desert scenario.
It is very plausible that the fundamental scale of unification is the Planck
scale, at which none of the known interactions can be neglected.
Many of the observations at low energies will arise from a
fundamental theory at the Planck scale.
The most developed Planck scale theories, to date, are superstring
theories. In this talk I discuss the construction of realistic
superstring models and their phenomenological implications.

Initially it was hoped that the uniqueness of the heterotic string [\HETE]
in ten dimensions would lead to a unique heterotic string theory in
four dimensions. However, soon thereafter it was realized that in four
dimensions there is a large number of consistent theories.
Viable models can be constructed by compactifying the extra dimensions
on a Calabi--Yau manifold [\CY] or on an orbifold [\DHVW].
Alternatively, one can
formulate consistent string theories by identifying the extra degrees
of freedom as an internal conformal field theory in the form of
free bosons [\NARAIN], free fermions [\FFF],
or as a product of minimal models [\GEPNER].
Thus the initial hope that consistency alone will determine
the string vacuum did not materialize.

Further progress can be made by pursuing a dual approach.
On the one hand we must study the theoretical aspects of superstring theory
and understand its fundamental principles at the
perturbative and nonperturbative levels. We then hope to learn how
the true string vacuum is selected. Alternatively, we may try to construct
realistic string models by imposing phenomenological constraints.
The realistic string models may then be used as a testing ground to test
our ideas on string theory and to study how Planck scale physics may
determine the parameters of the Standard Model.

Two approaches can be pursued to connect superstring theory with the
Standard Model. One is through a GUT model at an intermediate
energy scale [\REVAMP,\SUTHREE]. The second possibility is to derive the
Standard Model directly from superstring theory [\SSM--\SLM]. Due to proton
decay considerations the second possibility is preferred.
Consider the dimension four operators,
${\eta_1}{u_{L}^C}{d_{L}^C}{d_{L}^C}+{\eta_2}{d_{L}^C}QL$,
that exist in the most general supersymmetric Standard Model.
Unless ${\eta_1}$ and ${\eta_2}$ are highly suppressed,
these dimension four operators mediate rapid proton decay.
In the minimal supersymmetric Standard Model
one imposes a discrete symmetry, R parity, that forbids these terms.
In the context of superstring theories these discrete symmetries are
usually not present. If $B-L$ is gauged as in $SO(10)$, these dimension
four operators are forbidden by gauge invariance. However they may still
be induced from the nonrenormalizable terms,
$${\eta_1}({u_{L}^C}{d_{L}^C}{d_{L}^C}N_L^C)\Phi+
{\eta_2}({d_{L}^C}QLN_L^C)\Phi,\eqno(1)$$
where $\Phi$ is a string of $SO(10)$ singlets that fixes the string
selection rules and gets a VEV of $O(M_{Pl})$. $N_L^C$ is the Standard Model
singlet in the $16$ of $SO(10)$. It is seen that the ratio
${\langle{N_L^c}\rangle}/{M_{Pl}}$ controls the rate
of proton decay. Consequently, the VEV ${\langle{N_L^c}\rangle}$ has to be
suppressed. In superstring GUT models, that have been constructed to date,
${\langle{N_L^c}\rangle}$ is used to break the GUT symmetry because
there are no adjoint representations in the massless spectrum.
Next, consider proton decay from dimension five operators. Dimension five
operators are induced in SUSY GUT models by exchange of Higgsino color
triplets [\W]. Proton lifetime constraint requires that
Higgsino color multiplets are sufficiently heavy,
of the order of $10^{16}~GeV$.
Supersymmetric GUT models must admit some doublet--triplet splitting
mechanism, which satisfies these requirements. Although, such a mechanism
has been constructed in different supersymmetric GUT models, in general,
further assumptions have to be made on the matter content and
interactions of the supersymmetric GUT models. If the Standard Model
gauge group is obtained directly at the string level then we can construct
models in which the Higgsino color triplets are projected out from the massless
spectrum by the GSO projections. Thus, the proton lifetime considerations
motivate us to conjecture that in a realistic string model the Standard
Model nonabelian gauge group must be obtained directly at the string level.

In view of the large number of, a priori, possible string models, trying
to construct one realistic model may not seem very meaning full.
It is very plausible that models with some realistic features may
be constructed in different regions of the compactification space.
What would then tell us why one is preferred over the other.
However, not all the points in the compactification space are alike.
String theory exhibits a new kind of symmetry, usually referred to as
target space duality [\GPR],
which is a generalization of the $R\rightarrow{1/ R}$
duality in the case of $S^1$. At the self--dual point, $R_j={1/ R_j}$,
space--time symmetries are enhanced. For appropriate choices of the
background fields the space--time symmetries are maximally
enhanced. At the maximally symmetric point the internal
degrees of freedom that are needed to cancel the conformal anomaly
may be represented in terms of internal free fermions propagating on
the string world--sheet. It is not outrageous to assume that if
string theory has anything to do with nature the true string model
will be located near this highly symmetric point. Thus,
we are led to consider superstring standard--like models in the
free fermionic formulation.

\bigskip
\ll{\bf 2. Superstring standard--like models}

The superstring standard--like models are constructed in the free fermionic
formulation. In the free fermionic formulation [\FFF] of the heterotic string
in four dimensions all the world--sheet
degrees of freedom  required to cancel
the conformal anomaly are represented  in terms of free fermions
propagating on the string world--sheet.
For the left--movers  (world--sheet supersymmetric) one has the
usual space--time fields $X^\mu$, $\psi^\mu$, ($\mu=0,1,2,3$),
and in addition the following eighteen real free fermion fields:
$\chi^I,y^I,\omega^I$  $(I=1,\cdots,6)$, transforming as the adjoint
representation of $SU(2)^6$. The supercurrent
is given in terms of these fields as follows
$$T_F(z) = \psi^\mu\partial_zX_\mu + {\sum_{i=1}^6}\chi^iy^i\omega^i.$$
For the right movers we have ${\bar X}^\mu$ and 44 real free fermion fields:
${\bar\phi}^a$, $a=1,\cdots,44.$
Under parallel transport around a noncontractible loop the
fermionic states pick up a phase. A model in this construction
is defined by a set of  basis vectors of boundary conditions for all
world--sheet fermions. These basis vectors are constrained by the string
consistency requirements (e.g. modular invariance) and
completely determine the vacuum structure of
the model. The physical spectrum is obtained by applying the generalized
GSO projections. The low energy effective field theory is obtained
by S--matrix elements between external states. The Yukawa couplings
and higher order nonrenormalizable terms in the superpotential
are obtained by calculating correlators between vertex operators.
For a correlator to be nonvanishing all the symmetries of the model must
be conserved. Thus, the boundary condition vectors determine the
phenomenology of the models.

The first five vectors (including the vector {\bf 1}) in the basis
consist of the NAHE{\footnote*{This set was first
constructed by Nanopoulos, Antoniadis, Hagelin and Ellis  (NAHE)
in the construction
of  the flipped $SU(5)$.  {\it nahe}=pretty, in
Hebrew.}} set
$$\eqalignno{S&=({\underbrace{1,\cdots,1}_{{\psi^\mu},
{\chi^{1..6}}}},0,\cdots,0
\vert 0,\cdots,0).&(2a)\cr
b_1&=({\underbrace{1,\cdots\cdots\cdots,1}_
{{\psi^\mu},{\chi^{12}},y^{3,...,6},{\bar y}^{3,...,6}}},0,\cdots,0\vert
{\underbrace{1,\cdots,1}_{{\bar\psi}^{1,...,5},
{\bar\eta}^1}},0,\cdots,0).&(2b)\cr
b_2&=({\underbrace{1,\cdots\cdots\cdots\cdots\cdots,1}_
{{\psi^\mu},{\chi^{34}},{y^{1,2}},
{\omega^{5,6}},{{\bar y}^{1,2}}{{\bar\omega}^{5,6}}}}
,0,\cdots,0\vert {\underbrace{1,\cdots,1}_{{{\bar\psi}^{1,...,5}},
{\bar\eta}^2}}
,0,\cdots,0).&(2c)\cr
b_3&=({\underbrace{1,\cdots\cdots\cdots\cdots\cdots,1}_
{{\psi^\mu},{\chi^{56}},{\omega^{1,\cdots,4}},
{{\bar\omega}^{1,\cdots,4}}}},0,\cdots,0
\vert {\underbrace{1,\cdots,1}_{{\bar\psi}^{1,...,5},
{\bar\eta}^3}},0,\cdots,0).&(2d)\cr}$$
with the choice of generalized GSO projections
$c\left(\matrix{b_i\cr
                                    b_j\cr}\right)=
c\left(\matrix{b_i\cr
                                    S\cr}\right)=
c\left(\matrix{1\cr
                                    1\cr}\right)=-1,$
and the others given by modular invariance.

The gauge group after the NAHE set is $SO(10)\times
E_8\times SO(6)^3$ with $N=1$ space--time supersymmetry,
and 48 spinorial $16$ of $SO(10)$, sixteen from each sector
$b_1$, $b_2$ and $b_3$. The NAHE set divides the internal world--sheet
fermions in the following way: ${\bar\phi}^{1,\cdots,8}$ generate the
hidden $E_8$ gauge group, ${\bar\psi}^{1,\cdots,5}$ generate the $SO(10)$
gauge group, and $\{{\bar y}^{3,\cdots,6},{\bar\eta}^1\}$,
$\{{\bar y}^1,{\bar y}^2,{\bar\omega}^5,{\bar\omega}^6,{\bar\eta}^2\}$,
$\{{\bar\omega}^{1,\cdots,4},{\bar\eta}^3\}$ generate the three horizontal
$SO(6)^3$ symmetries. The left--moving $\{y,\omega\}$ states are divided
to $\{{y}^{3,\cdots,6}\}$,
$\{{y}^1,{y}^2,{\omega}^5,{\omega}^6\}$,
$\{{\omega}^{1,\cdots,4}\}$ and $\chi^{12}$, $\chi^{34}$, $\chi^{56}$
generate the left--moving $N=2$ world--sheet supersymmetry.

The internal fermionic states $\{y,\omega\vert{\bar y},{\bar\omega}\}$
correspond to the six left--moving and six right--moving compactified
dimensions in a geometric formulation. This correspondence is illustrated
by adding the vector
$X$ to the NAHE set, with periodic boundary conditions for the set
$({{\bar\psi}^{1,\cdots,5}},{{\bar\eta}^{1,2,3}})$ and antiperiodic
boundary conditions for all other world--sheet fermions.
This boundary condition vector extends the gauge symmetry to
$E_6\times U(1)^2\times E_8\times SO(4)^3$ with $N=1$ supersymmetry
and twenty-four chiral $27$ of $E_6$. The same model is generated in the
orbifold language [\DHVW]
by moding out an $SO(12)$ lattice by a $Z_2\times{Z_2}$
discrete symmetry with standard embedding. In the construction of
the standard--like models beyond the NAHE set, the assignment
of boundary conditions to the set of internal fermions
$\{y,\omega\vert{\bar y},{\bar\omega}\}$ determines many of the
properties of the low energy spectrum, such as the number of
generations, the presence of Higgs doublets, Yukawa couplings, etc.

In the realistic free fermionic models the boundary condition vector $X$
is replaced by the vector $2\gamma$ in which $\{{\bar\psi}^{1,\cdots,5},
{\bar\eta}^1,{\bar\eta}^2,{\bar\eta}^3,{\bar\phi}^{1,\cdots,4}\}$
are periodic and the remaining left-- and right--moving fermionic states
are antiperiodic. The set $\{1,S,2\gamma,\xi_2\}$ generates a model
with $N=4$ space--time supersymmetry and $SO(12)\times SO(16)\times SO(16)$
gauge group. The $b_1$ and $b_2$ twist are applied to reduce the number
of supersymmetries from $N=4$ to $N=1$ space--time supersymmetry.
The gauge group is broken to
$SO(4)^3\times U(1)^3\times SO(10)\times E_8$. The $U(1)$ combination
$U(1)=U(1)_1+U(1)_2+U(1)_3$ has a non--vanishing trace and the trace of the
two orthogonal combinations vanishes. The number of generations is still
24, eight from each sector $b_1$, $b_2$ and $b_3$
The chiral generations are now $16$ of $SO(10)$ from the sectors
$b_j$ $(j=1,2,3)$. The $10+1$ and the $E_6$ singlets from the sectors
$b_j+X$ are replaced by vectorial $16$ of the hidden $SO(16)$ gauge group
from the sectors $b_j+2\gamma$. As I will show below the structure of the
sector $b_j+2\gamma$ with respect to the sectors $b_j$ plays an important
role in the texture of fermion mass matrices.

The standard--like models are constructed by adding three additional
vectors to the NAHE set [\FNY,\EU,\TOP,\SLM]. The
the $SO(10)$ symmetry is broken in two stages,
first to $SO(6)\times SO(4)$ and next to $SU(3)\times SU(2)\times U(1)^2$.
One example is presented in the table, where only the boundary conditions
of the ``compactified space" are shown. In the gauge sector
$\alpha,\beta\{{{\bar\psi}^{1,\cdots,5}},
{{\bar\eta}^{1,2,3}},{\bar\phi}^{1,\cdots,8}\}=\{1^3,0^5,1^4,0^4\}$
and $\gamma\{{{\bar\psi}^{1,\cdots,5}},
{{\bar\eta}^{1,2,3}},{\bar\phi}^{1,\cdots,8}\}=\{{1\over2}^9,0,1^2,
{1\over2}^3,0\}$ break the symmetry to $SU(3)\times SU(2)\times
U(1)_{B-L}\times U(1)_{T_{3_R}}\times SU(5)_h\times SU(3)_h\times U(1)^2$.
The choice of generalized GSO coefficients is:
${c\left(\matrix{b_j\cr
                                    \alpha,\beta,\gamma\cr}\right)=
-c\left(\matrix{\alpha\cr
                                    1\cr}\right)=
c\left(\matrix{\alpha\cr
                                    \beta\cr}\right)=
-c\left(\matrix{\beta\cr
                                    1\cr}\right)}=$
${c\left(\matrix{\gamma\cr
                                    1,\alpha\cr}\right)=
-c\left(\matrix{\gamma\cr
                                    \beta\cr}\right)=
-1}$ (j=1,2,3), with the others specified by modular invariance and
space--time supersymmetry.

\bigskip
\ll{\bf 3.  The number of generations}

Free fermionic models with the NAHE set correspond to $Z_2\times Z_2$
orbifold at a special point in toroidal compactification space.
At this point the internal compactified dimensions can be represented
in terms of free world--sheet fermions. At this specific point the
symmetries due to the compactified dimensions are enhanced from $U(1)^6$
to $SO(12)$. The enhancement is due both to compactification at the self--dual
point $R_j=1/R_j$ and due to specific values of the background fields.
The structure of the $Z_2\times Z_2$ orbifold, with standard embedding,
at the specific point in compactification space is the root of the
realistic properties of the free fermionic models.
The first requirement from any superstring model is that the low
energy spectrum contains a net chirality of three generations.
A general $Z_2\times Z_2$ orbifold would not produce three generation
models. However, miraculously, at the most symmetric point in compactification
space, three generations are obtained very naturally. The reason
is that at this point the number of fixed points, in each twisted sector,
can be simultaneously reduced to one fixed point, from each twisted sector.
The three generations are then aligned along the three orthogonal complex
planes of the $Z_2\times Z_2$ orbifold. At the level of the NAHE set
each sector $b_1$, $b_2$ and $b_3$ has eight generations.
Three additional vectors are needed to reduce the
number of generations to one generation from each sector $b_1$, $b_2$
and $b_3$. Each generation has horizontal symmetries that constrain
the allowed interactions. Each generation has two gauged $U(1)$
symmetries $U(1)_{R_j}$ and $U(1)_{R_{j+3}}$. For every right--moving
$U(1)$ symmetry there is a corresponding left--moving global $U(1)$ symmetry
$U(1)_{L_j}$ and $U(1)_{L_{j+3}}$. Finally, each generation has two Ising
model operators that are obtained by pairing a left--moving real fermion
with a right--moving real fermion.
\vfill
\eject
\input tables.tex
\smallskip
{{\it Table 1.} A three generations ${SU(3)\times SU(2)\times U(1)^2}$
                model [\TOP].
\vskip .75mm
{\hfill
{\begintable
\  \ \|\
${y^3y^6}$,  ${y^4{\bar y}^4}$, ${y^5{\bar y}^5}$,
${{\bar y}^3{\bar y}^6}$
\ \|\ ${y^1\omega^6}$,  ${y^2{\bar y}^2}$,
${\omega^5{\bar\omega}^5}$,
${{\bar y}^1{\bar\omega}^6}$
\ \|\ ${\omega^1{\omega}^3}$,  ${\omega^2{\bar\omega}^2}$,
${\omega^4{\bar\omega}^4}$,  ${{\bar\omega}^1{\bar\omega}^3}$  \crthick
${\alpha}$ \|
1, ~~~1, ~~~~1, ~~~~0 \|
1, ~~~1, ~~~~1, ~~~~0 \|
1, ~~~1, ~~~~1, ~~~~0 \nr
${\beta}$ \|
0, ~~~1, ~~~~0, ~~~~1 \|
0, ~~~1, ~~~~0, ~~~~1 \|
1, ~~~0, ~~~~0, ~~~~0 \nr
${\gamma}$ \|
0, ~~~0, ~~~~1, ~~~~1 \|\
1, ~~~0, ~~~~0, ~~~~0 \|
0, ~~~1, ~~~~0, ~~~~1
\endtable}}
\bigskip
\ll{\bf 4. Higgs doublets }

Higgs doublets in the standard--like models are obtained from two
distinct sectors. The first type are obtained from the Neveu--Schwarz sector,
which produces three pairs of electroweak doublets
$\{h_1, h_2, h_3, {\bar h}_1, {\bar h}_2, {\bar h}_3\}$.
The Neveu--Schwarz sector corresponds to the untwisted sector of the
orbifold models.
Each pair of Higgs doublets can couple
at tree level only to the states from the sector $b_j$. This
results from the horizontal symmetries, $U(1)_j$,  $(1,2,3)$ and is a
reflection of the structure of the $Z_2\times Z_2$ twisting.
There is a
stringy doublet--triplet splitting mechanism that projects out the
color triplets and leaves the electroweak doublets in the spectrum.
Thus, the superstring standard--like models resolve the GUT hierarchy
problem. The second type of Higgs doublets are obtained from
the vector combination $b_1+b_2+\alpha+\beta$. The states in this
sector are obtained by acting on the vacuum with a single fermionic
oscillator and transform only under the observable sector.

In addition to electroweak doublets and color triplets the
Neveu--Schwarz sector and the sector $b_1+b_2+\alpha+\beta$
produce singlets of $SO(10)\times E_8$. These singlets play an
important role in the phenomenology of the superstring standard--like
models. The VEVs of the $SO(10)$ singlet fields in the massless
spectrum of the superstring models determine the light Higgs representations
and generate the fermion mass hierarchy.

The Neveu--Schwarz sector and the sector $b_1+b_2+\alpha+\beta$
produce four [\EU] or five [\TOP] pairs of electroweak doublets.
Several pairs receive heavy mass from the VEVs of Standard Model singlets
in the massless spectrum.
At the cubic level there are two pairs of electroweak doublets.
The light Higgs doublets are combinations of $(h_1,h_2,h_{45})$
and $({\bar h}_1,{\bar h}_2,{\bar h}_{45})$. The Higgs doublets
$h_3$ and ${\bar h}_3$ obtain a large mass from $SO(10)$ singlet VEVs.
This results from requiring F--flatness of the cubic level
superpotential [\NRT]. The absence of $h_3$
and ${\bar h}_3$ from the light eigenstates results in $b_3$ being
identified with the lightest generation.
At the nonrenormalizable level one additional pair receives a superheavy
mass and one pair remains light to give masses to the fermions at
the electroweak scale. Requiring F--flatness imposes that the light
Higgs representations are ${\bar h}_1$ or ${\bar h}_2$ and $h_{45}$.

\bigskip
\ll{\bf 5. The sectors $b_j+2\gamma$}

As mentioned above the realistic free fermionic models contain massless
states from the sectors $b_j+2\gamma$ $(j=1,2,3)$. These states arise due
to the $Z_2\times Z_2$ twisting on a gauge lattice with
$SO(16)\times SO(16)$ rather than $E_8\times E_8$. Thus, the realistic free
fermionic models correspond to $(2,0)$ rather than $(2,2)$ compactification.
The sectors $b_j+2\gamma$ produce the vectorial $16$ representation of the
hidden $SO(16)$ gauge group, decomposed under the final hidden gauge group.
The number of $16$ of the hidden $SO(16)$ is equal to the number of $16$ of
the observable $SO(10)$ gauge group. The horizontal charges in the sectors
$b_j+2\gamma$ are similar to the ones in the sectors $b_j$. The VEVs
of the states from the sectors $b_j+2\gamma$ are responsible
for generating texture zeroes in the fermion mass matrices.

The massless spectrum described until now results from the
$Z_2\times Z_2$ twist with standard embedding. Therefore, it generally
holds for all the free fermionic models that are based on
$Z_2\times Z_2$ orbifold with standard embedding.
In addition to the states from the sectors mentioned above there
are massless sectors that arise due to the sectors that correspond to Wilson
line breaking. The states in these sectors usually do not have the standard
$SO(10)$ embedding. In particular the weak hypercharge and $U(1)_{Z^\prime}$
charge usually differs from the standard $SO(10)$ assignment.

\bigskip
\ll{\bf 6. Top quark mass hierarchy }

Trilinear and nonrenormalizable
contributions to the superpotential are obtained by calculating correlators
between vertex operators [\KLN]
$$A_N\sim\langle V_1^fV_2^fV_3^b\cdot\cdot\cdot V_N^b\rangle, \eqno(3)$$
where $V_i^f$ $(V_i^b)$ are the fermionic (scalar)
components of the vertex operators. The non--vanishing terms are obtained by
applying the rules of Ref. [\KLN].
The cubic level Yukawa couplings for the quarks and leptons are
determined by the boundary conditions in the vector $\gamma$
according to the following rule [\SLM,\YUKAWA]
$$\eqalignno{\Delta_j&=
\vert\gamma(U(1)_{L_{j+3}})-\gamma(U(1)_{R_{j+3}})\vert=0,1
{\hskip 1cm}(j=1,2,3)&(4a)\cr
\Delta_j&=0\rightarrow d_jQ_jh_j+e_jL_jh_j;&(4b) \cr
\Delta_j&=1\rightarrow u_jQ_j{\bar h}_j+N_jL_j{\bar h}_j,&(4c)\cr}$$ where
$\gamma(U(1)_{R_{j+3}})$, $\gamma(U(1)_{L_{j+3}})$ are the boundary
conditions of the world--sheet fermionic currents that generate the
$U(1)_{R_{j+3}}$, $U(1)_{L_{j+3}}$ symmetries.

The superstring standard--like models contain an anomalous $U(1)$
gauge symmetry. The anomalous $U(1)$ generates a Fayet--Iliopoulos term
by the VEV of the dilaton field that breaks supersymmetry and destabilizes
the vacuum [\DSW].
Supersymmetry is restored by giving VEVs to Standard Model
singlets in the massless spectrum of the superstring models. However,
as the charge of these singlets must have $Q_A<0$ to cancel the anomalous
$U(1)$ D--term equation, in  many models a phenomenologically realistic
solution does not exist. In fact a very restricted class of standard--like
models with $\Delta_j=1$ for $j=1,2,3$, were found to admit a solution
to the F and D flatness constraints. Consequently, the only models that
were found to admit a solution are models which have tree level
Yukawa couplings only for $+{2\over3}$ charged quarks.

This result suggests an explanation for the top quark mass hierarchy
relative to the lighter quarks and leptons. At the cubic level only the
top quark gets a mass term and the mass terms for the lighter
quarks and leptons are obtained from nonrenormalizable terms.
To study this scenario we have to examine the nonrenormalizable
contributions to the doublet Higgs mass matrix and to the fermion mass
matrices [\NRT,\CKM].

At the cubic level there are two pairs of electroweak doublets.
At the nonrenormalizable level one additional pair receives a superheavy
mass and one pair remains light to give masses to the fermions at
the electroweak scale. Requiring F--flatness imposes that the light
Higgs representations are ${\bar h}_1$ or ${\bar h}_2$ and $h_{45}$.

The nonrenormalizable fermion mass terms of order $N$ are of the form
$cgf_if_jh\phi^{^{N-3}}$ or
$cgf_if_j{\bar h}\phi^{^{N-3}}$, where $c$ is a
calculable coefficient, $g$ is the gauge coupling at the unification
scale,  $f_i$, $f_j$ are the fermions from
the sectors $b_1$, $b_2$ and $b_3$, $h$ and ${\bar h}$ are the light
Higgs doublets, and $\phi^{N-3}$ is a string of Standard Model singlets
that get a VEV and produce a suppression factor
${({{\langle\phi\rangle}/{M}})^{^{N-3}}}$ relative to the cubic
level terms. Several scales contribute to the generalized VEVs. The
leading one is the scale of VEVs that are used to cancel
the ``anomalous'' $U(1)$
D--term equation. The next scale is generated by Hidden sector
condensates. Finally, there is a scale which is related to the breaking
of $U(1)_{Z^\prime}$, $\Lambda_{Z^\prime}$. Examination of the higher
order nonrenormalizable terms reveals that $\Lambda_{Z^\prime}$ has
to be suppressed relative to the other two scales.

At the cubic level only the top quark gets a nonvanishing mass term.
Therefore only the top quark mass is characterized by the electroweak
scale. The remaining quarks and leptons obtain their mass terms from
nonrenormalizable terms. The cubic and nonrenormalizable terms in the
superpotential are obtained by calculating correlators between the vertex
operators. The top quark Yukawa coupling is generically given by
$$g\sqrt2\eqno(5)$$
where $g$ is the gauge coupling at the unification scale. In
the model of Ref. [\TOP], bottom quark and tau lepton mass terms
are obtained at the quartic order,
$$W_4=\{{d_{L_1}^c}Q_1h_{45}^\prime\Phi_1+{e_{L_1}^c}L_1h_{45}^\prime\Phi_1+
{d_{L_2}^c}Q_2h_{45}^\prime{\bar\Phi}_2
+{e_{L_2}^c}L_2h_{45}^\prime{\bar\Phi}_2\}.\eqno(6)$$
The VEVs of $\Phi$ are obtained from the cancelation of the anomalous
D--term equation. The coefficient of the quartic order mass terms
were calculated by calculating the quartic order correlators and the
one dimensional integral was evaluated numerically. Thus after
inserting the VEV of ${\bar\Phi}_2$ the effective bottom quark and tau lepton
Yukawa couplings are given by [\TOP],
$$\lambda_b=\lambda_\tau=0.35g^3.\eqno(7)$$
They are suppressed relative to the top Yukawa by
$${{\lambda_b}\over{\lambda_t}}=
{{0.35g^3}\over{g\sqrt2}}\sim{1\over8}.\eqno(8)$$
To evaluate the top quark mass, the three Yukawa couplings are run
to the low energy scale by using the MSSM RGEs. The bottom mass is
then used to calculate $\tan\beta$ and the top quark mass is
found to be [\TOP],
$$m_t\sim175-180GeV.\eqno(9)$$
The fact that the top Yukawa is found near a fixed point suggests that this
is in fact a good prediction of the superstring standard--like models.
By varying $\lambda_t\sim0.5-1.5$ at the unification
scale, it is found that $\lambda_t$ is always $O(1)$ at the
electroweak scale.

\bigskip
\ll{\bf 7. Fermion mass matrices}

An analysis of fermion mass terms up to order $N=8$ revealed the general
texture of fermion mass matrices in these models. The sectors $b_1$
and $b_2$ produce the two heavy generations.
Their mass terms are suppressed by singlet VEVs that are used in the
cancellation of the anomalous $U(1)$ D--term equation.
The sector $b_3$ produces the lightest generation.
The diagonal
mass terms for the states from $b_3$ can only be generated by
VEVs that break $U(1)_{Z^\prime}$. This is due to the horizontal
$U(1)$ charges and because the Higgs pair $h_3$ and ${\bar h}_3$
necessarily gets a Planck scale mass [\NRT].
The suppression of the lightest generation mass terms is seen to be
a result of the structure of the vectors $\alpha$ and $\beta$
with respect to the sectors $b_1$, $b_2$ and $b_3$.
The mixing between the generations
is obtained from exchange of states from the sectors $b_j+2\gamma$. The
general texture of the fermion mass matrices in the superstring
standard--like models is of the following form,
$${M_U\sim\left(\matrix{\epsilon,a,b\cr
                    {\tilde a},A,c \cr
                    {\tilde b},{\tilde c},\lambda_t\cr}\right);{\hskip .2cm}
M_D\sim\left(\matrix{\epsilon,d,e\cr
                    {\tilde d},B,f \cr
                    {\tilde e},{\tilde f},C\cr}\right);{\hskip .2cm}
M_E\sim\left(\matrix{\epsilon,g,h\cr
                    {\tilde g},D,i \cr
                    {\tilde h},{\tilde i},E\cr}\right)},$$
where $\epsilon\sim({{\Lambda_{Z^\prime}}/{M}})^2$.
The diagonal terms in capital letters represent leading
terms that are suppressed by singlet VEVs, and
$\lambda_t=O(1)$. The mixing terms are generated by hidden sector states
from the sectors $b_j+2\gamma$ and are represented by small letters. They
are proportional to $({{\langle{TT}\rangle}/{M}^2})$.

\bigskip
\ll{\bf 8. Quark flavor mixing}

In Ref.
[\CKM] it was shown that if the states from the sectors $b_j+2\gamma$ obtain
VEVs in the application of the DSW mechanism, then a Cabibbo angle of the
correct order of magnitude can be obtained in the superstring standard--like
models. For one specific choice of singlet VEVs that solve the cubic
level F and D constraints the down mass matrix $M_D$ is given by
$$M_d\sim\left(\matrix
{&\epsilon
&{{V_2{\bar V}_3\Phi_{45}}\over{M^3}} &0\cr
&{{V_2{\bar V}_3\Phi_{45}\xi_1}\over{M^4}}
&{{{\bar\Phi}_2^-\xi_1}\over{M^2}} &0 \cr
&0 &0
&{{\Phi_1^+\xi_2}\over{M^2}}\cr}\right)v_2,\eqno(10)$$
where $v_2=\l h_{45} \r$ and we have used
${1\over2}g\sqrt{2\alpha^\prime}=\sqrt{8\pi}/M_{Pl}$, to define
$M\equiv M_{Pl}/2\sqrt{8\pi}\approx 1.2\times 10^{18}GeV$ [\KLN].
The undetermined VEVs of $\bar \Phi_{13}$ and $\xi_2$ are used to fix $m_b$ and
$m_s$ such that $\l \xi_1 \r \sim M$. We also take $tan \beta=v_1/v_2
\sim 1.5$.
Substituting the values of the VEVs above and
diagonalizing $M_D$ by a biunitary transformation we obtain the
Cabibbo mixing matrix
$$\vert V \vert\sim \left(\matrix {0.98&0.2&0 \cr
                                        0.2&0.98&0 \cr
                                        0&0&1 \cr } \right). \eqno(11)$$
Since the running from the scale $M$ down to the weak scale does not affect
the Cabibbo angle by much [\GLT], we conclude that realistic mixing of the
correct order of magnitude can be obtained in this scenario.
The analysis was extended to show that reasonable
values for the entire CKM matrix parameters can be obtained for appropriate
flat F and D solutions. For one specific solution the up and down quark mass
matrices take the form
$$M_u\sim\left(\matrix{&\epsilon
&{{V_3{\bar V}_2\Phi_{45}\bar \Phi_3^+}\over{M^4}} &0\cr
&{{V_3{\bar V}_2\Phi_{45}\bar \Phi_2^+}\over{M^4}}
&{{{\bar\Phi}_i^-\bar \Phi_i^+}\over{M^2}}
&{V_1{\bar V}_2\Phi_{45}\bar \Phi_2^+}\over{M^4} \cr
&0 &{V_1{\bar V}_2\Phi_{45}{\bar\Phi}_1^+}\over{M^4}
&1\cr}\right)v_1,\eqno(12)$$ and
$$M_d\sim\left(\matrix{&\epsilon
&{{V_3{\bar V}_2\Phi_{45}}\over{M^3}} &0\cr
&{{V_3{\bar V}_2\Phi_{45}\xi_1}\over{M^4}}
&{{{\bar\Phi}_2^-\xi_1}\over{M^2}} &{V_1{\bar V}_2\Phi_{45}\xi_i}\over{M^4} \cr
&0 &{V_1{\bar V}_2\Phi_{45}\xi_i}\over{M^4}
&{{\Phi_1^+\xi_2}\over{M^2}}\cr}\right)v_2,\eqno(13)$$
with $v_1$, $v_2$ and $M$ as before.
The up and down quark mass matrices are diagonalized
by bi--unitary transformations
$$\eqalignno{&U_LM_uU_R^\dagger=D_u\equiv{\rm diag}(m_u,m_c,m_t),&(14a)\cr
	     &D_LM_dD_R^\dagger=D_d\equiv{\rm diag}(m_d,m_s,m_b),&(14b)\cr}$$
with the CKM mixing matrix given by
$$V=U_LD_L^\dagger.\eqno(15)$$
The VEVs of $\xi_1$ and $\xi_2$ are fixed to be $\langle \xi_1 \rangle
\sim M/12$ and $\langle \xi_2 \rangle \sim M/4$ by the masses $m_s$ and $m_b$
respectively. Substituting the
VEVs and diagonalizing $M_u$ and $M_d$ by a bi--unitary transformation, we
obtain the mixing matrix
$$\vert V \vert\sim \left(\matrix {0.98&0.205&0.002 \cr
                                  0.205&0.98&0.012 \cr
                                 0.0004&0.012&0.99 \cr} \right). \eqno(16)$$
The texture and hierarchy of the mass terms in Eqs. (11--12) arise
due to the set of singlet VEVs in Eqs. (29).
The zeroes in the 13 and 31 entries of the mass matrices are protected to all
orders of nonrenormalizable terms. To obtain a non--vanishing
contribution to these entries either $V_1$ and ${\bar V}_3$
or $V_3$ and ${\bar V}_1$ must obtain a VEV simultaneously. Thus,
there is a residual horizontal symmetry that protects these vanishing
terms. The 11 entry in the mass matrices, e.g. the diagonal mass terms
for the lightest generation states, can only be obtained from VEVs
that break $U(1)_{Z^\prime}$ [\FM]. We assume that $U(1)_{Z^\prime}$
is broken at an intermediate energy scale that is suppressed relative
to the scale of scalar VEVs [\NRT]. In Ref. [\FHn] we showed that
$U(1)_{Z^\prime}$ is broken by hidden sector matter condensates
at $\Lambda_{Z^\prime}\leq10^{14}GeV$. Consequently,
we have taken $\epsilon\leq(\Lambda_{Z^\prime}/M)^2\sim10^{-8}$.

Texture zeroes in the fermion mass matrices are
obtained if the VEVs of some states from the sectors $b_j+2\gamma$ vanish.
These texture zeroes are protected by the symmetries of the string models
to all order of nonrenormalizable terms [\CKM]. For example in the above mass
matrices the 13 and 31 vanish because $\{V_1,V_3\}$ get a VEV but ${\bar V}_1$
and ${\bar V}_3$ do not. Therefore these mass matrix terms cannot be formed
because they would not be invariant under all the string symmetries.
Other textures are possible for other choices of VEVs for the states from the
sectors $b_j+2\gamma$.
\vfill
\eject
\bigskip
\ll{\bf 9. Neutrino masses}

A seesaw type neutrino mass matrix can be constructed from analysis of
nonrenormalizable terms and for specific choices of singlet VEVs [\FHn].
The neutrino seesaw mass matrix takes the general form
The neutrino mass matrix therefore takes the following form for each generation
in the basis $(\nu_L, N^C, \Phi)$
$$\left( \matrix{0&km_u&0 \cr
          km_u&0&m_\chi \cr
          0&m_\chi&m_\phi \cr} \right)~, \eqno(17)$$
with $m_\chi \sim  \left( \Lambda_{Z^\prime}\over M \right)^3\left({\langle
\phi \rangle }\over M \right)^n M$ and $m_\phi \sim \left(\Lambda_{Z^\prime}
\over M \right)^4 \left(\langle \phi \rangle \over M \right)^m M$.
$n$ and $m$ are the orders at which the terms are obtained.
The mass eigenstates are mainly $\nu$, $N$ and $\phi$ with a small mixing and
with the eigenvalues
$$m_{\nu} \sim m_\phi \left({{k m_u} \over m_{\chi}}\right)^2
\qquad m_N,M_{\phi} \sim m_{\chi} \eqno(18)$$
The constant $k$ gives the effects of Yukawa coupling renormalization.
The seesaw scale $m_\chi$
is determined by the $U(1)_{Z^\prime}$ breaking scale
and by the order at which the nonrenormalizable seesaw terms are obtained.
In Ref. [\FHn] the $U(1)_{Z^\prime}\sim10^{14}~GeV$
breaking scale was obtained from condensates of the hidden $SU(5)$ gauge
group with nontrivial $U(1)_{Z^\prime}$ charges.
The order of nonrenormalizable terms that contribute to the seesaw
terms in the neutrino mass matrix depends highly on the choice of flat
flat directions. Neutrino masses that are in agreement with experimental
constraints can be obtained. A novel feature of the superstring
seesaw mechanism is that although the $U(1)_{Z^\prime}$ breaking scale
may be large (e.g. $\Lambda_{Z^\prime} \approx 10^
{14} GeV$) the effective see-saw scale can be much smaller.

\bigskip
\ll{\bf 9. Gauge coupling unification}

While LEP results indicate that the gauge coupling in the minimal
supersymmetric Standard Model unify at $10^{16}GeV$, superstring
theory predicts that the unification scale is at $10^{18}GeV$.
The superstring standard--like models may resolve this
problem due to the existence of color triplets and electroweak doublets
from exotic sectors that arise from the additional vectors $\alpha$,
$\beta$ and $\gamma$. These exotic states carry fractional charges and
do not fit into standard $SO(10)$ representations. Therefore, they
contribute less to the evolution of the $U(1)_Y$ beta function than
standard $SO(10)$ multiplets. For example in Ref. [\GCU] representations
with the following beta function coefficients, in a
${SU(3)}\times {SU(2)\times {U(1)_Y}}$ basis, were found
$$b_{D_1,{\bar D}_1,D_2,{\bar D}_2}={\left(\matrix{{1\over2}\cr
                      0\cr
                     {1\over5}\cr}\right)};
  b_{D_3,{\bar D}_3}={\left(\matrix{{1\over2}\cr
                      0\cr
                     {1\over{20}}\cr}\right)};
  b_{\ell,{\bar\ell}}={\left(\matrix{0\cr
                      {1\over2}\cr
                         0\cr}\right)}.$$
The standard--like models predict
$\sin^2\theta_W={3/8}$ at the unification scale due to the
embedding of the weak hypercharge in $SO(10)$. In Ref. [\GCU],
I showed that provided that the additional exotic color triplets and
electroweak doublets exist at the appropriate scales, the
scale of gauge coupling unification is pushed to $10^{18}GeV$, with the
correct value of $\sin^2\theta_W$ at low energies.

\bigskip
\ll{\bf 11. Hierarchical SUSY breaking}

In Ref. [\SUSYX]
we address the following question: Given a supersymmetric string
vacuum at the Planck scale, is it possible to obtain hierarchical
supersymmetry breaking in the observable sector? A supersymmetric
string vacuum is obtained by finding solutions to the cubic level
F and D constraints. We take a gauge coupling in agreement with
gauge coupling unification, thus taking a fixed value for the dilaton VEV.
We then investigate the role of nonrenormalizable terms and strong
hidden sector dynamics. The hidden sector contains two non--Abelian
hidden gauge groups, $SU(5)\times SU(3)$, with matter in vector--like
representations. The hidden $SU(3)$ group is broken near the Planck scale.
We analyze the dynamics of the hidden $SU(5)$ group.
The  $SU(5)$ hidden matter mass matrix is given by
$${\cal M}=\left(\matrix{ 0   & C_1   & 0   \cr
			  B_1 & A_2   & C_2 \cr
			  0   & C_3   & A_1   \cr  }\right)~,\eqno(19)$$
where $A,B,C$ arise from nonrenormalizable terms of orders $N=5,8,7$
respectively and are given by
$$\eqalignno{A_1&= {{\l\Phi_{45}{\bar\Phi}_1^-\xi_2\r}\over{M^2}},
\qquad \qquad A_2= {{\l\Phi_{45}\Phi_2^+\xi_1\r}\over{M^2}}
		   						,&(20a,b)\cr
B_1&= {{\l V_3{\bar V_2}\Phi_{45}\Phi_{45}{\bar\Phi}_{13}\xi_1\r}\over
{M^5}},&(20c)\cr
C_1&= {{\l V_3{\bar V_2}\Phi_{45}\Phi_{45}{\bar\Phi}_{13}\r}\over{M^4}},~~
C_2 = {{\l V_1{\bar V_2}\Phi_{45}\Phi_{45}\xi_1\r}\over{M^4}},&(20d,e)\cr
C_3&= {{\l V_1{\bar V_2}\Phi_{45}\Phi_{45}\xi_2\r}\over{M^4}}~~.
&(20f)}$$
Taking generically $\l\phi\r\sim{gM}/4\pi\sim M/10$ we obtain
$A_i \sim 10^{15}~GeV$, $B_i \sim 10^{12}~GeV$, and $C_i\sim 10^{13}~GeV$.
{}From Eqs. (19--20) we observe that to insure a nonsingular
hidden matter mass matrix, we must require $C_1\ne0$ and $B_1\ne0$.
This imposes ${\bar V_3}\ne0$ and $V_2\ne0$. Thus, the nonvanishing VEVs
that generates the Cabibbo mixing also guarantee the stability of the
supersymmetric vacuum.
The gaugino and matter
condensates are given by the well known expressions for supersymmetric $SU(N)$
with matter in $N+{\bar N}$ representations [\SQCD],
$$\eqalignno{
{1\over{32\pi^2}}\vev{\lambda\lambda}
&=\Lambda^3\left(det{{\cal M}\over\Lambda}\right)^{1/N},&(21a)\cr
\Pi_{ij}=\vev{{\bar T_i}T_j}&={1\over{32\pi^2}}\vev
{\lambda\lambda}{{\cal M}_{ij}}^{-1},&(21b)\cr}$$
where $\l\lambda\lambda\r$, $\cal M$ and
$\Lambda$ are the hidden gaugino condensate, the hidden matter mass matrix
and the $SU(5)$ condensation scale, respectively.
Modular invariant generalization of Eqs. (20a,b) for the string case were
derived in Ref. [\LT].
The nonrenormalizable terms can be put in modular invariant form
by following the procedure outlined in Ref. [\KLNI].
Approximating the Dedekind $\eta$ function by $\eta({\hat T})\approx
e^{-\pi {\hat T}/12}(1-e^{-2\pi {\hat T}})$
we verified that the calculation using the modular invariant expression
from Ref. [\LT] (with $\l {\hat T}\r\approx M$) differ from the results using
Eq. (20), by at most an order of magnitude.
The hidden $SU(5)$ matter mass matrix is nonsingular for specific F and D
flat solutions. In Ref. [\SUSYX] a specific cubic level F and D flat
solution was found. The gravitino mass due to the gaugino and matter
condensates was estimated to be of the order $1-10~TeV$.
The new aspect of our scenario for supersymmetry breaking is the
following. As long as only states from the Neveu--Schwarz sector
or the sector $b_1+b_2+\alpha+\beta$ receive VEVs in the application
of the DSW mechanism then one can find exact flat directions at the
cubic level of the superpotential. These flat directions will be exact
and will not be spoiled by nonrenormalizable terms. The states from the
Neveu--Schwarz sector and the sector $b_1+b_2+\alpha+\beta$ correspond
to untwisted and twisted moduli.
However, once some hidden sector matter states obtain a nonvanishing
VEV, the cubic level flat directions are no longer exact. Supersymmetry
is broken by the inclusion of nonrenormalizable terms. Hidden sector
strong dynamics at an intermediate scale may then be responsible for
generating the hierarchy in the usual fashion.

\bigskip
\ll{\bf 12. conclusion}

The Standard Model is in agreement with all current experiments.
Furthermore, present day experiments seem to
support the big desert scenario and the
notion of unification. The Planck scale is the ultimate scale
of unification at which none of the known interactions can be
ignored. Many properties of the Standard Model will arise from
the fundamental Planck scale theory.
Superstring theory stands out as the only known theory
that can consistently unify gravity with the gauge interactions.
The heterotic string is the only string theory that can produce
realistic phenomenology. Its consistency requires twenty--six critical
dimensions in the bosonic sector and ten critical dimensions in the
supersymmetric sector. In the bosonic sector sixteen degrees of
freedom are compactified on a flat torus and produce the observable
and hidden gauge degrees of freedom. Six degrees of freedom from the
bosonic sector, combined with six degrees of freedom from the
supersymmetric sector, are compactified on a Calabi--Yau manifold
or on an orbifold. String theory exhibits a new kind of symmetry:
``target space duality''. At the self--dual point, the compactified
degrees of freedom can be represented in terms of free world--sheet
fermions. At this point space--time symmetries are maximally enhanced.
The most realistic superstring models constructed to date were
constructed at this point in the compactification space.
The underlying structure of the
$Z_2\times Z_2$ orbifold at the free fermionic point in toroidal
compactification space is the origin of the realistic nature of free fermionic
models. We believe that if string unification is relevant in nature,
then the underlying structure of the
$Z_2\times Z_2$ orbifold at the free fermionic point in toroidal
compactification space will be intrinsic to the eventual ``true'' heterotic
string model. Thus, it makes sense, in our opinion, to try to build
realistic models specifically at this point in the huge
compactification space.

The superstring standard--like models contain in their
massless spectrum all the necessary states to obtain realistic low energy
phenomenology. They resolve the problems of proton decay through
dimension four and five operators that are endemic to other
superstring and GUT models. The existence of only three generations with
standard $SO(10)$ embedding is understood to arise naturally from
$Z_2\times Z_2$ twisting at the free fermionic point in compactification
space. Better understanding of the correspondence with other superstring
formulations will provide further insight into the realistic properties
of these models. In this context it is especially interesting to try to
understand the significance of the self--dual point in the compactification
space. Finally, the free fermionic standard-like models
provide a highly constrained and phenomenologically realistic laboratory
to study how the Planck scale may determine the parameters of the
Standard Model.

\bigskip
\ll{\bf 12. Acknowledgments}

This work is supported by an SSC fellowship.
Part of the work described in this talk was done in collaboration
with Edi Halyo.

\smallskip
\baselineskip=12pt
\refout
\end